\documentstyle[12pt]{article}
\def\1{\mbox{I\hspace{-.15em}1}}

\def\b{\begin{equation}}
\def\e{\end{equation}}
\def\bee{\begin{enumerate}}
\def\eee{\end{enumerate}}

\title{Alternative Inflationary Scenario\\ Due to Compact Extra Dimension}
\author{E. Yusofi$^{1}$\thanks{e-mail:
eyph2009@gmail.com}  and M. Mohsenzadeh$^{2}$}
\date{\today}
\begin{document}

\maketitle {\it\centerline{\it $^{1}$ Department of
Physics,Islamic Azad University- Amol Branch,Iran \\ P.O.Box 678,
Amol, Mazandaran} \centerline{\it $^{2}$ Department of
Physics,Islamic Azad University- Qom Branch, Iran, Qom, IRAN}}

\begin{abstract}
The main goal of this paper is to give an alternative
interpretation of space-like and time-like extra dimensions as a
primary factor for inflation in the early universe. We introduce
the 5-dimensional perfect fluid and compare the energy-momentum
tensor for the bulk scalar field with space-like and time-like
extra dimensions. It is shown, that additional dimensions can
imply to negative pressure in the slow roll regime in the early
higher-dimensional world.
\end{abstract}

\vspace{0.5cm} {\it Proposed PACS numbers}: ???


\section{Introduction}
The 5-dimensional Warped geometry theory is a braneworld theory
developed by Physicists, Lisa Randall and Ramam Sundrum, while
trying to solve the hierarchy problem of the Standard Model[1-4]. They
consider one extra "non-factorizable" dimension. The metric is
assumed to be \b
ds^{2}=g_{ab}dx^{a}dx^{b}=\eta_{\mu\nu}e^{-2kry}dx^{\mu}dx^{\nu}+r^{2}dy^{2}\e
which $a,b=(0,1,2,3,4)$ and Greek indices $\mu,\nu=(0,1,2,3)$, where refer to
the four observable dimensions, and also $y$ signifies the
coordinate on the additional dimension of length $r$ , and the
factor $e^{-2kry}$ is called Warp factor. The metric tensor in
this model given by \b g_{ab}=
\textit{diag}(e^{-2A(y)},-e^{-2A(y)},-e^{-2A(y)},-e^{-2A(y)},r^{2})\e
Which $ e^{-2A(y)} $ is called generalized Warp factor[2].
We have for space-like extra dimensions (SLED), $ r^{2}=-1 $ and
for time-like extra dimensions (TLED), $ r^{2}=+1 $. Very often,
the standard tenet in dealing with higher-dimensional theories is
to consider space-like extra dimensions. But time-like extra
dimensions have been disregarded due to the serious conflicts with
causality and unitarity[5-7].

Hear it was assumed to have two branes. One at $ y=0 $, called the
Planckbrane ( where gravity is a relatively strong force; also
called Gravitybrane), and one at $ y=r\pi $, called the Tevbrane (
our home with the Standard Model particles; also called
Weakbrane). As the neam already tells all standard fields are
assume to live on the second brane. By Warping any lagrangian mass
parameter which is naturally $ \approx{M_{PL}} $, will appear to
us on the SM brane to be $ \approx{TeV} $. Thus by considering
Warped extra dimensions one is able to solve the hierarchy
problem. This means that transition from 4D world to 5D world,
exponentially shrink size and grow mass and energy[1,8-10].

In this paper we have considered opposite process of above, {\it
i.e.} transition from 5D world to the 4D of it. By consideration
of symmetry, it can be expected exponentially expansions of size
occur in this transition. Therefore, we can foretell the
alternative inflationary scenario for 5D world before of the
standard inflation in 4D world. Thus, the outline of paper is as
follows: In section 2, the energy-momentum tensor for 4D metric is
recalled briefly. In section 3, we introduce the 5D perfect fluid
and compare of the energy-momentum tensor for the balk scalar
field with space-like and time-like extra dimensions. Brief
conclusions are given in final section.

\section{The Energy-momentum Tensor for 4D metric}
In this section, we cosider a single scalar field, called $
inflaton field $ during the period of inflation. The lagrangian of
the inflaton field[11-15], $ \phi $, is \b
L=\frac{1}{2}g_{\mu\nu}\partial^{\mu}\phi\partial^{\nu}\phi-V(\phi)
\e Where \b g_{\mu\nu}=
\textit{diag}(a^{2}(\eta),-a^{2}(\eta),-a^{2}(\eta),-a^{2}(\eta))\e
is the FRW metric with the conformal time coordinate $ \eta $, by
the definition $ dt=a(\eta)d\eta $. Also $ a(\eta) $ is the scalar
factor which depends only on time.

The action for the inflaton field is  \b
S=\int{d^{4}x\sqrt{-g}L}=\int{d^{4}x\sqrt{-g}}[\frac{1}{2}g_{\mu\nu}\partial^{\mu}\phi\partial^{\nu}\phi-V(\phi)]
\e The energy-momentum tensor for the inflaton field is given by[16],
\b T_{\mu\nu}=-g_{\mu\nu}L+\partial_{\mu}\phi\partial_{\nu}\phi\e
For any perfect fluid, this tensor given by \b T^{\mu}_{\nu}=
\textit{diag}(\rho,-P,-P,-P)\e Hear $\rho$ and $P$ are the density
and pressure of the perfect fluid respectively. Remember that
"perfect" can be taken to means "isotropic in its rest frame".
This in turn means that $T^{\mu}_{\nu}$ is diagonal - there is
flux of any component of momentum in an orthogonal direction.
Furthermore, the nonzero space-like components must all be equal,
{\it i.e.} $ T^{1}_{1}=T^{2}_{2}=T^{3}_{3}$[16]. Because $
T_{\mu\nu}=g_{\mu\beta}T^{\beta}_{\nu} $, one obtains \b
T_{\mu\nu}= \textit{diag}(\rho
a^{2}(\eta),Pa^{2}(\eta),Pa^{2}(\eta),Pa^{2}(\eta))\e Considering
the inflaton field as a homogeneous perfect fluid, its
energy-momentum tensor in (6) can be written in components as \b
T_{00}=[\frac{1}{2}(\frac{\partial\phi}{\partial\eta})^{2}+V(\phi)a^{2}(\eta)]
\e \b T_{0i}=0 \e \b
T_{ij}=[\frac{1}{2}(\frac{\partial\phi}{\partial\eta})^{2}-V(\phi)a^{2}(\eta)]\delta_{ij}
\e Comparing the results with (8), the energy density and the
pressure of the inflaton field are \b
\rho=[\frac{1}{2a^{2}(\eta)}(\frac{\partial\phi}{\partial\eta})^{2}+V(\phi)]
\e \b
P=[\frac{1}{2a^{2}(\eta)}(\frac{\partial\phi}{\partial\eta})^{2}-V(\phi)]
\e It can be seen that when the potential energy of the inflaton
field is larger than its Kinetic energy ( slow-roll condition),
the negative pressure appears. This state, {\it i.e.} $ P=-\rho $
is very important in order to have inflation $(\ddot{a}>0)$.

\section{The Energy-momentum Tensor for 5D metric}
The lagrangian of the scalar field,$\phi $, in 5-dimensional world
is \b L=\frac{1}{2}g_{ab}\partial^{a}\phi\partial^{b}\phi-V(\phi)
\e Which the field $\phi=\phi(y) $, is a balk scalar field which
respect to only the extra dimension coordinate $y$.

The general form of action is  \b
S=\int{d^{5}x\sqrt{-g}L}=\int{d^{4}xdy\sqrt{-g}}[\frac{1}{2}g_{ab}\partial^{a}\phi\partial^{b}\phi-V(\phi)]
\e And the energy-momentum tensor is given by \b
T_{ab}=-g_{ab}L+\partial_{a}\phi\partial_{b}\phi\e Because of the
homogeneity, the nonzero space-like and time-like components must
all be equal. Hence for 5-dimensional perfect fluid, we introduce
two shapes of tensor, for SLED \b T^{a}_{b}=
\textit{diag}(\rho,-P,-P,-P,-P^{5})\e And for TLED \b T^{a}_{b}=
\textit{diag}(\rho,-P,-P,-P,\rho^{5})\e Because of homogeneity
condition $ \rho^{5}=\rho $ and $ P^{5}=P $ are the density and
pressure of the 5D perfect fluid respectively.

Because $ T_{ab}=g_{a\beta}T^{\beta}_{b} $, one obtains from
(1),(17) and (18) \b T_{ab}=
\textit{diag}(e^{-2A(y)}\rho,e^{-2A(y)}P,e^{-2A(y)}P,e^{-2A(y)}P,P^{5})\e
and \b T_{ab}=
\textit{diag}(e^{-2A(y)}\rho,e^{-2A(y)}P,e^{-2A(y)}P,e^{-2A(y)}P,\rho^{5})\e
Considering the bulk scalar field as a homogeneous perfect fluid,
its energy-momentum tensor in (16) can be written in components as
\b T_{00}=e^{-2A}[\frac{1}{2}\phi'^{2}+V(\phi)] \e \b
T_{ij}=-T_{00}=-e^{-2A}[\frac{1}{2}\phi'^{2}+V(\phi)]\delta_{ij}
\e \b T_{44}=[\frac{1}{2}\phi'^{2}-V(\phi)] \e  for SLED[3]
and \b T_{00}=e^{-2A}[-\frac{1}{2}\phi'^{2}+V(\phi)] \e
\b
T_{ij}=-T_{00}=-e^{-2A}[-\frac{1}{2}\phi'^{2}+V(\phi)]\delta_{ij}
\e  \b T_{44}=[\frac{1}{2}\phi'^{2}+V(\phi)] \e for TLED. Hear,
the prime denotes the derivative with respect to only fifth
coordinate $y$. Comparing the results (21),(22),(23) with (19) and
(24),(25),(26) with (20), the consequent energy density and the
pressure for the balk scalar field for SLED are \b
\rho=[\frac{1}{2}\phi'^{2}+V(\phi)] \e \b
P=-[\frac{1}{2}\phi'^{2}+V(\phi)] \e \b
P^{5}=[\frac{1}{2}\phi'^{2}-V(\phi)] \e and for TLED are \b
\rho=[-\frac{1}{2}\phi'^{2}+V(\phi)] \e \b
P=[\frac{1}{2}\phi'^{2}-V(\phi)] \e \b
\rho^{5}=[\frac{1}{2}\phi'^{2}+V(\phi)] \e

In the 5-dimensional Warped geometry model, any particles and
fields moving from the Planckbrane to the Tevbrane in the bulk
would be growing, becoming lighter, and moving more slowly through
time . Distance and time expand near the Tevbrane, and also mass
and energy shrink near it[8-10]. The slow moving of
the scalar field in the bulk can be similar to the slow rolling
condition for inflaton field in standard 4D inflationary model. Therefore with this condition the potential energy of this field is larger
than its gradient(Kinetic) energy $ (\frac{1}{2}\phi'^{2}) $,
the negative pressure appears. These states, {\it i.e.} $
P^{5}=\rho $ and $ P=-\rho^{5} $ is very impotant in order to have
inflation in 5D early universe $(\ddot{a}>0)$.

For 5D inflation, the fifth components of energy-momentum tensor,
$\rho^{5} $ and $P^{5} $ is more important than other associated
components in (19) and (20). Because on the Planckbrane, strings
would be $ 10^{-33}cm $ in size, but on the Tevbrane, they'd be  $
10^{-17}cm $ . In fact, this makes the energy and mass scale for
the Planckbrane based on about  $ 10^{16}TeV $ . Therefor, the
fifth components of pressure and density in compact size situate
on the exited mode of KK model and have an unstable state. This
unstable and exited mode in 5D cosmic system soon can be undergoes
a transition to a stable zero mode in 4D cosmic system, a
phenomenon can be known as spontaneous symmetry breaking.
Consequently, scalar field set in 4D universe and additional
dimensions disappearance.

\section{Conclusion}
This paper provide an alternative mechanism for another inflationary
scenario of early universe vs. 4D standard inflation. This mechanism produce of transition from unstable state of 5D world with planck scale of energy to the 4D world with  TeV scale of energy. It
has been shown consideration 5D perfect fluid with SLED and
TLED, can leads to negative pressure for energy-momentum tensor of the bulk scalar field. This condition is very important in order to have inflationary expansions for higher-dimensional world.

\noindent {\bf{Acknowlegements}}:
This work has been supported by
the Islamic Azad University-Ayatollah Amoli Branch, Amol,
Mazandaran, Iran.

\end{document}